# Quality Assessment of the 20m SPOT DEM in Nigeria

[1]Nwilo, P.C., [1*]Onyegbula, J.C., [1]Okolie, C.J., [1]Daramola, O.E. [1]Abolaji, O.E. and [2]Arungwa, I.D.

[1]Department of Surveying and Geoinformatics, University of Lagos, Nigeria.
[2]Department of Surveying and Geoinformatics, Abia State University, Nigeria.

*Corresponding author email: johansononyegbula20@gmail.com

**Abstract**
The 20m SPOT DEM (Digital Elevation Model) was acquired by the Office of the Surveyor-General of the Federation (OSGoF) in Nigeria for use in topographic mapping. A localized assessment of the DEM is needed to validate its stated accuracies as it is well known that the vertical accuracy of global DEMs varies in different landscape contexts. Hence, this study assessed the quality of the DEM in variable land cover types by comparing its heights against 780 Ground Control Points (GCPs) in Lagos State and the Federal Capital Territory (FCT) of Nigeria. The pattern of distribution of the height differences was analysed using spatial autocorrelation analysis, including other accuracy metrics employed. In the general accuracy assessment, the DEM yielded root mean square errors of 2.33m in Lagos and 3.69m in the FCT. It was shown that heights over the bare lands were the most accurately represented on the DEM while heights over built-up areas are the least accurate at both locations. Despite this, the SPOT DEM accuracies in the varying land cover types surpassed its stated global accuracy. Also, Pearson's correlation analysis revealed no indication that error distribution over the landscape was a function of slope and/or aspect. While the spatial distribution of low and high height differences in Lagos State showed clustering, the converse was the case in the FCT. Following this assessment, it is recommended that the country should extend the application scope of the DEM in order to exploit its utility to the maximum.

**FUNDING**
This study did not receive any specific funding from the government, public or private sector.

**CONFLICT OF INTEREST DISCLOSURE**
The authors declare no conflict of interest.

**DATA AVAILABILITY STATEMENT**
The data that supports the findings of this study are available in Figshare at (https://figshare.com/articles/dataset/Quality_Assessment_of_the_20m_SPOT_DEM_in_Nigeria/13337042) for download. The raw data was provided by the Office of the Surveyor General of the Federation (OSGOF), Interspatial Technologies Ltd, Lagos State Surveyor General's

Office and the Federal Capital Development Authority (FCDA), FCT. Direct requests for these materials may be made to the provider as indicated in the Acknowledgements.

**CODE AVAILABILITY**
Not applicable.

**ETHICS APPROVAL**
Not applicable.

**CONSENT TO PARTICIPATE**
Not applicable.

**CONSENT FOR PUBLICATION**
The authors are fully aware and consent to the publication of this manuscript.

**1. INTRODUCTION**

Digital Elevation Models (DEMs) usually consist of elevation values represented in a pixel or cell array over an area (Fisher and Tate, 2006). Oftentimes, DEMs are represented with emphasis on the bare or natural surface, excluding all forms of vegetation, man-made features and other elements that describe topographic surfaces including skeleton, curvature and ridges (Podobnikar, 2008). Conventional methods of generating DEMs employ methods such as tacheometry, levelling and Global Navigation Satellite Systems (GNSS) surveys (Ravibabu and Jain, 2008). Other methods include aerial based photogrammetric methods using airplanes, Unmanned Aerial Vehicles (UAVs) and satellites (Croneborg et al., 2015). DEMs can also be derived from pre-existing topographic maps (Nwilo et al., 2017a) which contain contours, although interpolation is frequently required to estimate heights in between known elevations. Usually, space-based platforms such as satellites are the most common remote sensing methods employed for large-scale acquisition of data for DEMs today. Satellite-derived DEMs have proven very useful in modelling terrains of regions for various forms of geospatial analysis and other applications including hydrological modelling, terrain suitability for agriculture, slope and aspect analysis, town planning and numerous other applications (Al-Yami, 2014). The quality and reliability of DEMs acquired through satellite remote sensing is a subject of interest for researchers.

SPOT (Satellites Pour L'Observation de la Terre) Earth Observing Satellites is a commercial high-resolution optical imaging satellite system run by the French Space Agency in collaboration with the Belgian Scientific, Technical and Cultural Services (SSTC) and the Swedish National Space Board. It was launched to acquire high-resolution imagery of the Earth's surface for applications in environmental and resource monitoring, climatology, human activities, cartographic purposes, geospatial analysis and other applications. Several SPOT missions (SPOT 1 to SPOT 7) have been launched since 1986 (Nwilo et al., 2017b), acquiring data at varying spatial resolutions. One of the mission objectives of SPOT 5 was to offer high-resolution imagery for use in generating a DEM with an accuracy of 10m (EO Portal Directory, 2020). The 20-metre resolution DEM from SPOT 5 which was acquired with the High Resolution Stereoscopic (HRS) sensor is referenced to the Earth Gravity Model 1996 (EGM 96). The DEM has been assessed to a vertical accuracy of 10-20m (Baudoin et al., 2004; Li and Gruen, 2004; Reinartz et al., 2004; Massera et al., 2012; GISAT, 2019). The development of the HRS instrument was done by a French subsidiary of the European Aeronautic Defence and Space Company (EADS) Astrium with sponsorship by the National Centre for Space Studies in France (CNES) and SPOT Image (EO Portal Directory, 2020). The vast spatial coverage and improved geometric qualities of data acquired from SPOT missions (Rosengren and Willen, 2004), makes the SPOT DEM a useful dataset for modelling the terrain of diverse regions. SPOT DEMs are useful in topographic mapping of regions, town planning, generation of contours, and flood vulnerability assessment (Nwilo et al., 2017b).

With the increased usage of DEMs, a considerable amount of research has been carried out into the uncertainties or errors associated with the heights obtained from these models (Lidberg et al., 2017). Certain errors are bound to occur during the stages of acquisition to final processing of satellite-derived DEMs, leading to discrepancies between the DEM heights and the actual heights of the terrain (Olusina and Okolie, 2018). Artificial/man-made structures and trees pose another challenge to the accuracy of satellite-derived DEMs, and are difficult/nearly impossible to ignore. They sometimes mask the true height of the terrain by reflecting incident electromagnetic (E-M) radiation from the sensors, and consequently, have their heights representing the elevations of the terrain in DEMs. This is corroborated by the findings of Arungwa et al. (2018) and Nwilo et al. (2017a). Nwilo et al. (2017a) investigated the effects of above-ground landscape offsets on the 30-metre DEM from SRTM in three study sites located in Lagos and Ogun States of South-West Nigeria. The results showed a strong positive correlation was observed between the SRTM DEM and reference DEM with the highest correlation in bare lands ($R^2$ - 0.98) and lowest in wetland forests ($R^2$ – 0.85). Subsequent measures have to be taken to reduce these above-ground offsets to the barest minimum, as they are not representative properties required in a bare-earth DEM.

Closely linked to the accuracies of DEMs is the quality of their terrain derivatives. Terrain derivatives are topographic properties derived from DEMs and used to describe and analyze the terrain beyond the mere digital representation in a three-dimensional (3D) format which conventional DEMs would do. Some of these derivatives include first order derivatives such as slope and aspect, and others such as contours, flow direction, and hill shade. Slope and aspect are among some of the most important DEM derivatives used in natural resources spatial databases (Bolstad and Stowe, 1994). Slope refers to the rate of change of the surface, calculated from the pixel values or from the points on a triangulated irregular network (Maune, 2011); a high value for the slope indicates a steep terrain, while a lower value indicates a gentle or mildly undulating terrain. The aspect of a slope refers to the direction of a plane or surface measured in degrees – from 0 to 360 degrees, clockwise from the north. It indicates the direction a surface is facing which is useful for determining the amount of sunshine and wind received on that surface. The accuracy of a DEM and its derivatives are of crucial importance since the errors in the DEM will be propagated through the spatial analysis based on the data derived from it (Bolstad et al., 1994 in Wang, 1998). Hence, to effectively assess the quality of these DEMs, the relationship between the DEMs and their derivatives needs to be established.

The processing and manipulation of these DEMs and their derivatives is commonly carried out with Geographical Information Systems (GIS).

Previous studies on quality and accuracy of DEMs include the works of Gorokhovich and Voustianiouk (2006), Yastikli et al. (2006), Tighe and Chamberlain (2009), Hirt et al. (2010), Hengl and Reuter (2011), Rexer and Hirt (2014), and Arungwa et al. (2018). Also, the assessment of the relationship between DEM quality and terrain derivatives has received reasonable attention. For example, in the analysis of DEM accuracy in relation to slope, Gorokhovich and Voustianiouk (2006) assessed the accuracy of SRTM v4.1-based elevations with that of two independent datasets collected with the same Global Positioning System (GPS) in the Catskills Mountains (New York, USA) and in Phuket (Thailand). The study noted a strong correlation of the error values with slope and aspect. The analysis revealed significant decrease in accuracy when measurements were performed on terrain characterized by slope values greater than 10°. In another study, Miliaresis (2008) evaluated the effects of land cover on the aspect/slope accuracy dependence of the SRTM-1 Elevation Data for the Humboldt Range in the north-west portion of Nevada, USA. The SRTM elevations were compared with bare-earth elevations from the US National Land Cover Dataset (NLCD) and the US National Elevation Dataset (NED). The decomposition of elevation differences on the basis of aspect and slope terrain classes identified an over-estimation of elevation by the SRTM instrument along the east, north-east and north directions (negative elevation difference decreasing linearly with slope). Conversely, there was an under-estimation evident towards the west, south-west and south directions (positive elevation difference increasing with slope). Other studies on the relationship between DEM quality and terrain derivatives include Fisher (1991), Lee et al. (1992), Zhou and Liu (2004), A-Xing et al. (2008), and Qiming and Xuejun (2008).

In 2012, the Office of the Surveyor General of the Federation (OSGoF) in Nigeria acquired the 20m SPOT DEM for use in topographic mapping. This study presents an assessment of the SPOT DEM over parts of Nigeria in Lagos and the Federal Capital Territory (FCT). The assessment also considers the variation of the accuracy in different land cover types, and investigates the relationship between the DEM accuracy and terrain derivatives within the same study area. This is to advance the understanding of the reliability of the DEM for topographic mapping in Nigeria, especially the accuracy of its height attributes.

## 2. MATERIALS AND METHODS
### 2.1 Study Area

The study area includes Lagos State and the Federal Capital Territory (FCT) of Nigeria. Lagos State has a low-lying terrain and is located in the south-western part of Nigeria. It is one of the 36 states of the country and the smallest state by land mass with a total area of about 3,577.28km$^2$, 2797.72km$^2$ of land and 779.56km$^2$ of water (BudgIT, 2018) – constituting about 0.4% of the total land area of Nigeria. It is located between longitudes $2°41'15'' - 4°22'00''E$ and latitudes $6°22'20'' - 6°43'20''N$; consisting of 20 Local Government Areas (LGAs) and 37 Local Council Development Areas (LCDAs). The mangrove swamp forest and freshwater swamp forest are very dominant in Lagos State, with a drainage system of lagoons and waterways constituting about 22% of the state. The Lekki Lagoon and Lagos Lagoon are the major water bodies. Lagos is characterised by lowlands, gently increasing northwards with the presence of beaches, barrier islands, ports for transportation and international trade (Olusina and Okolie, 2018). The FCT is centrally located in Nigeria. It is geographically located between longitudes $6°45'50'' - 7°45'20''E$ and latitudes $8°26'10'' - 9°27'20''N$, with 6 area councils comprising its territory. It is a planned city and its geography is defined by Aso Rock, a 400m monolith. Topographically, the FCT is typified by a gently undulating terrain and riverine depressions (IPA, 1979 in Ojigi, 2006). It has rich soil for agriculture and is characterized by highlands and an undulating terrain which acts as a modulating influence for a clement weather (FCDA, 2020).

## 2.2 Datasets

The 20m SPOT DEM v1.0 (in *.tif format) was acquired from the Office of the Surveyor General of the Federation (OSGoF) Nigeria. Reference Ground Control Points (GCPs) for Lagos State were acquired from the Lagos State Surveyor General's Office. GCPs for FCT were acquired from the Department of Survey and Mapping, Federal Capital Development Authority and field observations were done to densify the second order controls. Liu et al. (2007) have listed among the factors that affect the accuracy of DEMs, the density and distribution of the source data. This also applies to the number and distribution of ground points which influences the estimation of DEM accuracy. For this study, a total number of 780 GCPs were acquired, 556 for Lagos and 224 for the FCT. Furthermore, Landsat 7 imageries were acquired for year 2002 to coincide with the year of acquisition of the SPOT DEM. Landsat imageries are downloadable from the Earth explorer website of the United States Geological Survey (USGS) - http://earthexplorer.usgs.gov. The path/row numbers of the Landsat imageries acquired are as follows: Lagos (191/055, 191/056 and 190/056) and FCT (189/053, 189/054). The GCP elevations were harmonised to the orthometric height system to match the height system of the SPOT DEM. Figure 1 presents a map of Nigeria showing the spatial distribution of the GCPs over the SPOT DEM coverage of the study area.

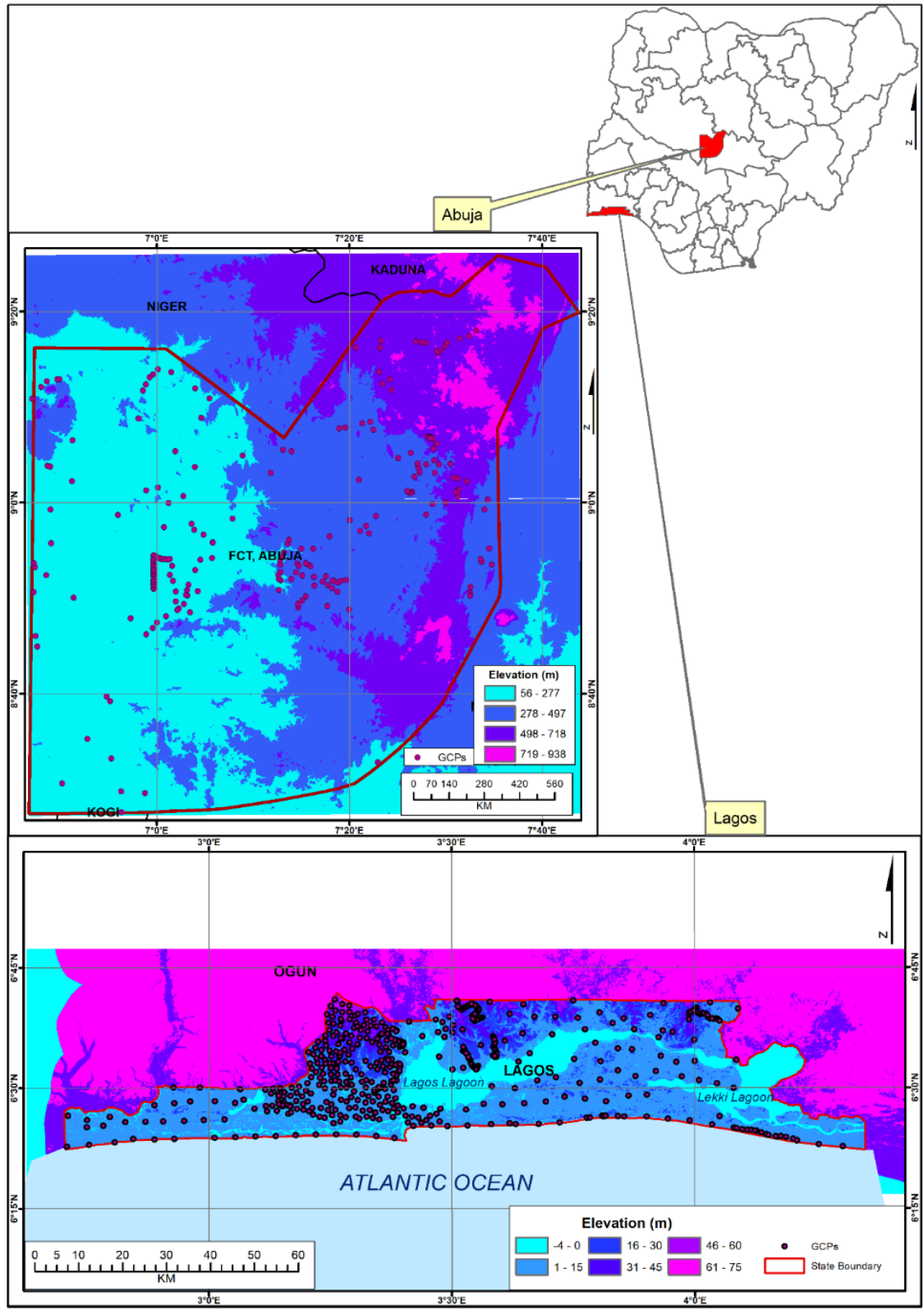

Figure 1: Map showing the location of the study areas, the SPOT DEM coverage and spatial distribution of the Ground Control Points

## 2.3 Extraction of Land Cover and Terrain Derivatives

Within ENVI software, false colour composites were generated from the Landsat 7 bands in the following order – Band 5, Band 4, Band 3. In the image interpretation, the following information classes were identified – Lagos (built-up area, bare land, forest land, wetland, and water body) and FCT (built-up area, bare land, forest land, agricultural land, water body and rocks). The composites were then subjected to supervised classification using the parallelepiped technique. Thereafter, the information classes were converted to shapefile format for further editing. Terrain derivatives (slope and aspect) were extracted directly from the DEM using the slope and aspect tools in ArcGIS 3D Analyst.

## 2.4 Extraction of Coincident Heights

The next step taken was the extraction of heights from the SPOT DEM at points coincident with the GCPs. This operation is necessary for a proper accuracy assessment of the heights extracted from the SPOT DEM relative to the GCPs, which for the purpose of this research, served as the reference dataset. This extraction was done using the 'Extract values to points' tool in the ArcGIS Spatial Analyst. In the extraction, the points were separated and grouped into different categories based on the intersecting land cover types with the exception of water bodies. After extraction, the heights were extracted from the shapefile .dbf attachment and copied to a Microsoft Excel worksheet.

## 2.5 Data Exploration

In the general descriptive analysis, data points intersecting with water bodies and unclassified areas were excluded. Also, a few points located on rock outcrops in FCT were deemed to be very sparse and excluded. This reduced the number of data points to 538 for Lagos and 221 for FCT. The next step considered an exploration of the data to screen for outliers prior to the accuracy assessment. This was done using the Tukey's method of outlier detection which defines outliers as values greater than $Q3 + 1.5 * IQR$ and values less than $Q1 - 1.5 * IQR$, where $Q1$, $Q3$ and $IQR$ are the lower quartile, upper quartile, and inter-quartile range respectively (Crawley, 2005). The run of Tukey's method returned some outliers in the data which were subsequently removed. In addition, data points with values below zero or those intersecting with water bodies were also eliminated. This further reduced the selection to 497 points for Lagos and 185 points for FCT.

## 2.6 Quantitative Analysis

The height differences ($\Delta H$) between the DEM and GCP points were calculated as follows:

$$\Delta H = H_{SPOT} - H_{GCP} \qquad (1)$$

Where,

$H_{GCP}$ = height from GCP

$H_{SPOT}$ = height from SPOT DEM

Using a combination of Microsoft Excel and Statistical Package for the Social Sciences (SPSS) software, the following accuracy parameters were computed for the height differences in the different land cover types: standard deviation (SD) and root mean square error (RMSE). The formula for SD follows from Spiegel and Stephens (1999).

$$SD = \sqrt{\sum_{i=1}^{n} \frac{(\Delta H_i - \underline{\Delta H})^2}{n-1}} \qquad (2)$$

Similarly, the formula for RMSE is given by Chai and Draxler (2014).

$$RMSE = \sqrt{\frac{1}{n}\sum_{i=1}^{n} \Delta H_i^2} \qquad (3)$$

Where,

$n$ = number of points

$\underline{\Delta H}$ = mean of the height differences

To assess the spatial pattern and distribution of the height differences, the Spatial Autocorrelation (Global Moran's I) analysis was carried out in the ArcGIS environment. The Spatial Autocorrelation analysis is a spatial statistical tool that determines the spatial distribution of a feature based on the feature values (Mitchell, 2005). In the analysis, the following parameters are computed; the Moran's I Index value, the z-score and the p-value, to evaluate the significance of the Index. According to Mitchell (2005), the Moran's I statistic for spatial autocorrelation is given as

$$I = \frac{n}{S_0} \frac{\sum_{i=1}^{n} \sum_{j=1}^{n} w_{i,j} z_i z_j}{\sum_{i=1}^{n} z_i^2} \qquad (4)$$

Where $z_i$ is the deviation of an attribute for feature $i$ from its mean ($x_i = \hat{X}$), $w_{i,j}$ is the spatial weight between features $i$ and $j$, $n$ is equal to the total number of features, and $S_0$ is the aggregate of all spatial weights.

$$S_0 = \sum_{i=1}^{n} \sum_{j=1}^{n} w_{i,j} \qquad (5)$$

The z score for the statistic is calculated as;

$$z_I = \frac{I - E[I]}{\sqrt{V[I]}} \qquad (6)$$

Where:

$$E[I] = \frac{-1}{(n-1)} \qquad (7)$$

$$V[I] = E[I^2] - E[I]^2 \qquad (8)$$

The relationship between elevations and terrain derivatives such as slope and aspect were evaluated using the Pearson's correlation coefficient (r). In the analysis, correlation was deemed significant at the 0.01 level (2-tailed). Going further, a one-way analysis of variance (ANOVA) was conducted to test for the presence of significant variations of the height differences ($\Delta H$) in different land cover types. One-way ANOVA is used to test for significance in variations between certain properties of 3 or more independent means, and makes use of the F-formula (Fischer ratio). The Fischer ratio for one-way ANOVA follows from Devore (2012).

For this analysis, the null hypothesis (H$_0$) used was set as $\underline{\Delta H}_1 = \underline{\Delta H}_2 = \underline{\Delta H}_3 = \underline{\Delta H}_4$. That is, the mean of $\Delta H$ for the various land cover types are the same or do not vary significantly. The alternative hypothesis (H$_1$) would imply the mean height differences vary significantly. A 0.05 level of significance was used for the test for comparison of acceptability of H$_0$.

## 3. RESULTS AND DISCUSSION
### 3.1 Heights and Land Cover Distribution

Figures 2 and 3 present the land cover maps for Lagos State and FCT respectively. The areal distribution of land cover for Lagos State is as follows: built-up areas (671.50km$^2$), bare lands (147.05km$^2$), wetlands (926.37km$^2$), forest lands (1,156.14km$^2$) and water bodies (774.67km$^2$). The distribution for FCT shows that built-up areas account for 508.54km$^2$ while bare lands cover 1,119.29km$^2$ of the total area. Other features in FCT include agricultural lands (4,998.16km$^2$), rocks (54.68km$^2$), forest lands (960.20km$^2$) and water bodies (26.17km$^2$). For the height analysis, all vegetation features were considered as a single class. Vegetation incorporates wetlands and forest lands in Lagos; and agricultural lands and forest lands in the FCT. Tables 1 and 2 present the descriptive statistics of heights at DEM and GCP coincident points in Lagos State and FCT respectively. In Lagos, SPOT DEM heights ranged from ~0 − 63m and in FCT, the range was from 74 – 737m. Mean heights of SPOT DEM in Lagos were derived as follows: bare land (10.17m), built-up area (17.66m), and vegetation (16.12m). In FCT, the mean heights were derived as follows: bare land (266m), built-up area (399.51m), and vegetation (320.95m). The mean heights in the generally low-lying terrain of Lagos State is least for bare land, which is quite expected as it is characterised by the presence of little or no features which can yield false (increased) heights. This mean height for the DEM increases in areas of vegetation cover which is expected as the presence of tall vegetation is expected to provide significant offset for reflected satellite signals, which can yield terrain overestimations. A similar explanation holds for built-up areas, where the presence of tall buildings provides the same masked height effect as the vegetation. The mean heights in FCT are relatively high due to the presence of highlands in its undulating terrain. The FCT landscape is profiled by isolated highlands, rolling hills and gaps with low dissected plains (NTDC, 1997 in Ojigi, 2006).

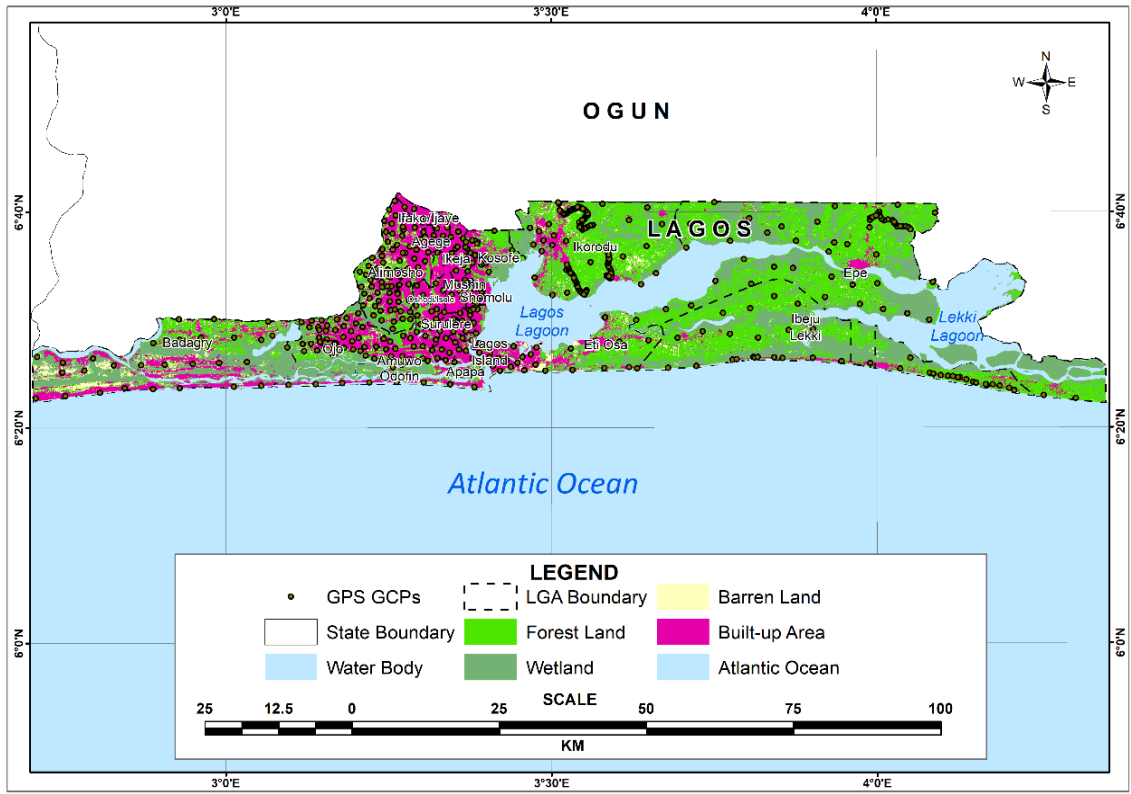

Figure 2: Land cover map of Lagos State showing the GCPs

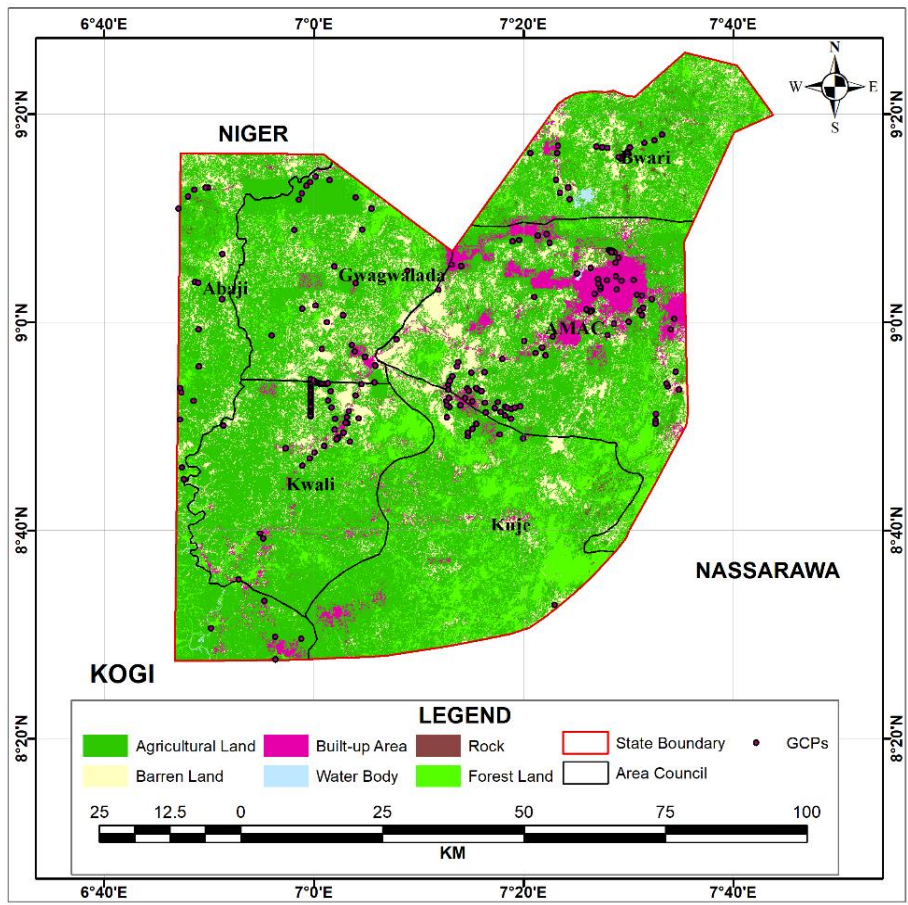

Figure 3: Land cover map of FCT showing the GCPs

Table 1: Descriptive statistics of heights at Lagos

| Statistic | Bare land | | Built-up area | | Vegetation | |
|---|---|---|---|---|---|---|
| | $H_{GCP}$ (m) | $H_{SPOT}$ (m) | $H_{GCP}$ (m) | $H_{SPOT}$ (m) | $H_{GCP}$ (m) | $H_{SPOT}$ (m) |
| Count | | 29 | | 292 | | 217 |
| Min. | 1.122 | 1 | 1.331 | -1 | 0.861 | -2 |
| Max. | 45.140 | 47 | 63.871 | 63 | 49.197 | 52 |
| Range | 44.018 | 46 | 62.541 | 64 | 48.337 | 54 |
| Mean | 9.760 | 10.172 | 16.634 | 17.661 | 15.703 | 16.124 |

Table 2: Descriptive statistics of heights at FCT

| Statistic | Bare land | | Built-up area | | Vegetation | |
|---|---|---|---|---|---|---|
| | $H_{GCP}$ (m) | $H_{SPOT}$ (m) | $H_{GCP}$ (m) | $H_{SPOT}$ (m) | $H_{GCP}$ (m) | $H_{SPOT}$ (m) |
| Count | | 45 | | 49 | | 127 |
| Min. | 108.939 | 110 | 139.452 | 137 | 65.254 | 74 |
| Max. | 680.575 | 685 | 596.331 | 589 | 748.335 | 737 |
| Range | 571.636 | 575 | 456.879 | 452 | 683.081 | 663 |
| Mean | 265.771 | 266 | 397.813 | 399.510 | 320.814 | 320.953 |

## 3.1 Accuracy Assessment

Tables 3 and 4 present the SDs and RMSEs of the height differences in Lagos and FCT respectively. In the general accuracy assessment, the DEM yielded RMSEs of 2.33m in Lagos and 3.69m in the FCT. This is a significant improvement to what was reported by Nwilo et al. (2020) and can be attributed to the exclusion of data outliers in the present study. The mean height difference in the different land cover types are as follows: bare land (Lagos: 0.12m; FCT: 1.46m), built-up area (Lagos: 0.91m; FCT: 2.84m), and vegetation (Lagos: 0.08m; FCT: 1.65m). Overall mean height differences are 0.54m and 1.88m for Lagos and FCT respectively. The SDs and RMSEs observed in the land cover types in Lagos are as follows: bare land (SD: 2.27m, RMSE: 2.23m), built-up area (SD: 2.22m, RMSE: 2.40m) and vegetation (SD: 2.24m; RMSE: 2.24m). The highest accuracy of the DEM in Lagos was in areas covered by bare lands with the least RMSE of 2.23m. Expectedly for bare land, the range of height differences is the least at both locations. In the FCT, the SDs and RMSEs in the different land cover types were derived as follows: bare land (SD: 2.03m, RMSE: 2.48m), built-up area (SD: 3.78m, RMSE: 4.69m) and vegetation (SD: 3.24m, RMSE: 3.62m). This shows that heights over the bare lands are the most accurately represented on the DEM while heights over built-up areas are the least accurate at both locations. The nature of the earth's surface in the built-up areas is rough and this affects the reflection of the signals from the satellite sensors (Pidwirny, 2006). Also, Lagos is densely populated with buildings and many of its commercial districts have tall buildings thereby increasing height error. The higher RMSE in areas of vegetation cover is explained by the inability of the sensor to sufficiently penetrate forested areas to sample the terrain. Forests and tree canopies pose as obstructions to the terrain height sampling by the satellite. Over the years, urban expansion in Lagos State and the FCT has led to increase in the total floor space of buildings in both residential areas and commercial business districts (Mahmoud et al. 2016;

Wang and Maduako, 2018). FCT has a hilly terrain with highlands in excess of 700m unlike Lagos that is generally low-lying. According to Aguilar et al. (2005), terrain morphology has an effect on DEM accuracy. The increased height errors in built-up areas and vegetation-covered areas is not uncommon with satellite DEMs. For example, height errors are also observed in Synthetic Aperture Radar DEMs due to double bounce scattering in urban areas (Delgado Blasco et al., 2020) and in vegetation-covered areas (Townsend, 2001; Schlaffer et al., 2015; Tsyganskaya et al., 2016).

Table 3: S.D and RMSE of height differences at Lagos

| Statistic | Bare land | Built-up area | Vegetation | Total |
|---|---|---|---|---|
| Count | 28 | 275 | 194 | 497 |
| Max (-ve) | -5.54 | -5.01 | -5.81 | -5.81 |
| Max (+ve) | 5.75 | 6.83 | 6.54 | 6.83 |
| Range | 11.29 | 11.83 | 12.35 | 12.64 |
| Mean | 0.12 | 0.91 | 0.08 | 0.54 |
| S.D | 2.27 | 2.22 | 2.24 | 2.26 |
| RMSE | 2.23 | 2.40 | 2.24 | 2.33 |

Table 4: S.D and RMSE of height differences at FCT

| Statistic | Bare land | Built-up area | Vegetation | Total |
|---|---|---|---|---|
| Count | 40 | 42 | 103 | 185 |
| Max (-ve) | -3.22 | -6.65 | -6.98 | -6.98 |
| Max (+ve) | 6.33 | 9.74 | 10.12 | 10.12 |
| Range | 9.55 | 16.39 | 17.10 | 17.10 |
| Mean | 1.46 | 2.84 | 1.65 | 1.88 |
| S.D | 2.03 | 3.78 | 3.24 | 3.19 |
| RMSE | 2.48 | 4.69 | 3.62 | 3.69 |

Figures 4 and 5 show histograms of height differences at Lagos and FCT respectively. Figures 4 (a-d) appear to be in harmony with summary statistics in Table 3 for Lagos. It indicates a normal distribution in the height differences of the DEM where the mean tends towards the value of the mode. The figures are great indicators of minimal influence of systematic error in the height estimates over Lagos. The values of the mean errors for Lagos (bare land: 0.12m; built-up area: 0.91m; vegetation: 0.08m; total: 0.54m) corroborates this fact. Given the fact that histograms are primary indicators of the presence of systematic errors, it is clear that from figures 5, there may be the presence of systematic error in the DEM in the region of FCT. As seen in Table 4, the mean error in all cases departs from zero (bare land: 1.46m; built-up area: 2.84m; vegetation: 1.65m; total: 1.88m).

Figures 6 and 7 show plots of height differences ($\Delta H$) against heights from SPOT DEM ($H_{SPOT}$) at Lagos and the FCT respectively. In Lagos, it is clear that height differences tend to be evenly distributed around the zero line. It can also be inferred that errors do not tend to increase in magnitude and frequency towards higher altitudes. On the contrary, and in previous studies of other DEMs within the area of interest and other areas of Nigeria, errors in DEMs have been observed to increase in magnitude in areas of higher altitude (Arungwa et al., 2018; Chigbu et al., 2019). One consistent feature in the FCT (Figure 7) is that most of the height differences are positive. (i.e. above the zero line). This indicates a general overestimation of the topography of the area of FCT. In addition to this is the fact that a large proportion of height differences are occurring in high altitude areas.

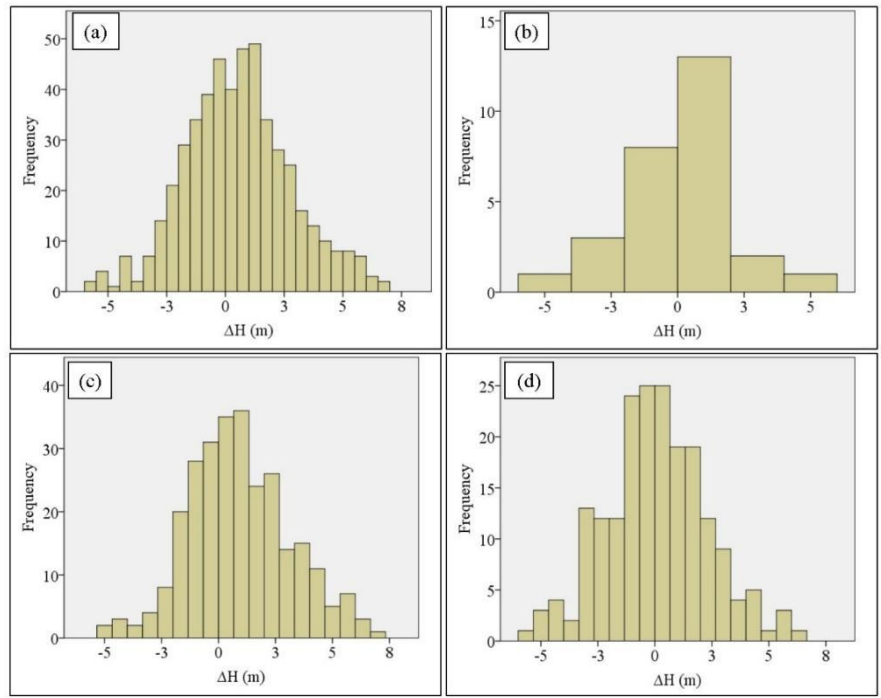

Figure 4: Histogram of height differences at Lagos (a) All points (b) Bare land (c) Built-up area (d) Vegetation

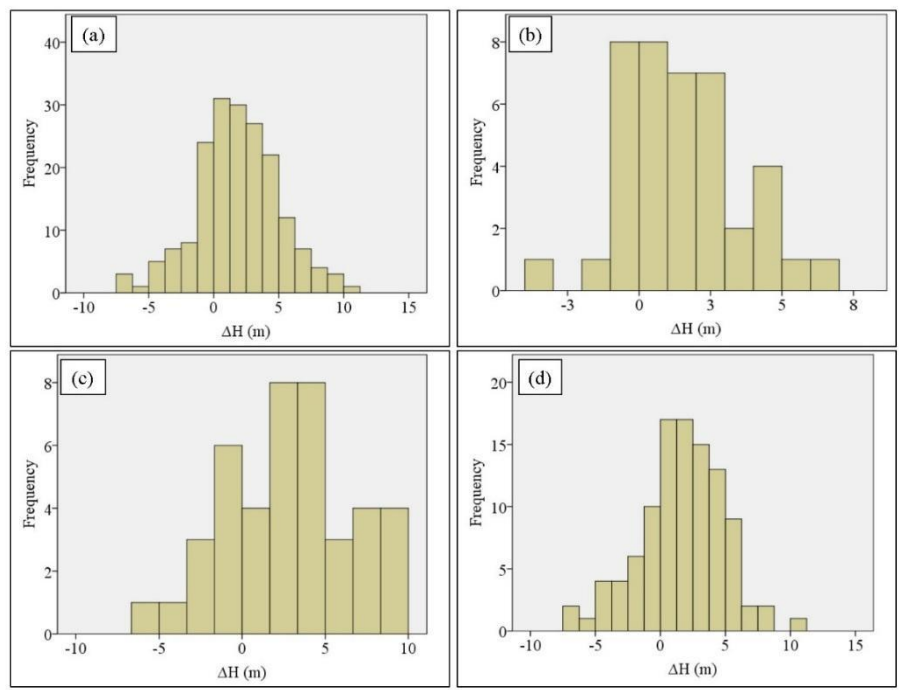

Figure 5: Histogram of height differences at FCT (a) All points (b) Bare land (c) Built-up area (d) Vegetation

*Nwilo P.C. et al. Link to final published version - https://rdcu.be/cMDQM*

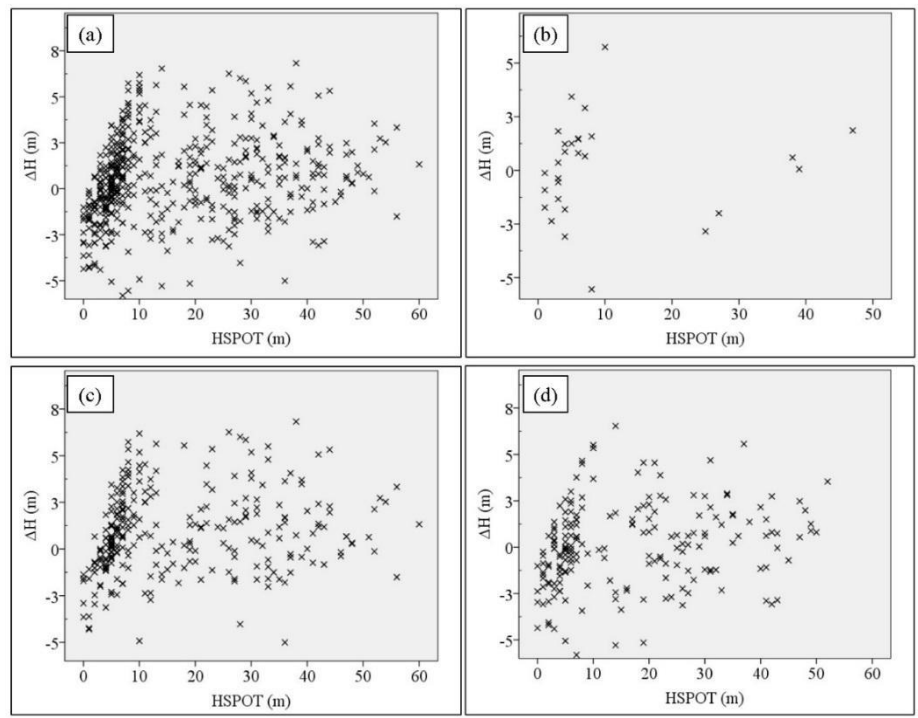

figure 6: Plots of height differences (Δ*H*) against heights from SPOT DEM (H$_{SPOT}$) at Lagos
(a) All points (b) Bare land (c) Built-up area (d) Vegetation

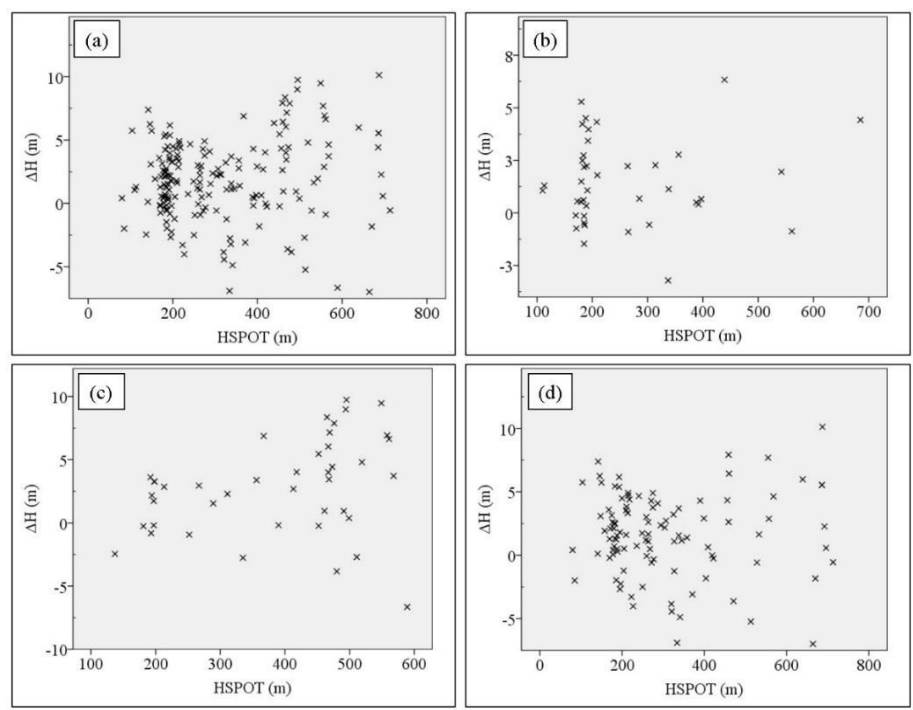

Figure 7: Plots of height differences (Δ*H*) against heights from SPOT DEM (H$_{SPOT}$) at FCT
(a) All points (b) Bare land (c) Built-up area (d) Vegetation

Spatial Autocorrelation (Global Moran's I) analysis was further carried out to determine the spatial pattern of the height differences. For Lagos, the height differences showed a clustered distribution with the Moran's I index of 0.108, z-score of 4.703 and p-value of 0.000. The lower p-value than 0.05 and the positive z-score show that the spatial distribution of high or low

**14**

height differences is significantly clustered. For FCT, the spatial distribution of the height differences yielded a Moran's I index of 0.698, z-score, 1.454 and p-value of 0.146. The higher p-value than 0.05 and the positive z-score show that the spatial distribution of high or low height differences in FCT does not exhibit clustering.

As a further check on the reliability of the DEM over the varied landscape, Table 5 presents the analysis of variance in the height differences at Lagos and FCT respectively. While there are significant differences in the height differences in the various land cover types within Lagos State, the differences in height are insignificant in the various land cover types in FCT. The significant differences in Lagos State are explained by the complexity and heterogeneity in its land cover with built-up areas and vegetation competing for space and occurring side by side in many areas. As such, satellite signals impact the above-ground obstructions at widely varying levels. Conversely, the FCT benefits from a better physical planning structure with many built-up areas existing separately from the vast forests and surrounding vegetation. However, available evidence suggests that there has been a rapid growth in satellite towns in the FCT (Mallo and Obasanya, 2011; Aliyu, 2016). The FCT is at a much higher elevation but over 70% of its land is covered by vegetation that reflect the satellite signals at above-ground levels that are near-uniform over vast areas with homogenous forest types in certain areas.

Table 5: Analysis of variance in height differences

| | | Sum of Squares | df | Mean Square | F | Sig. |
|---|---|---|---|---|---|---|
| Lagos | Between Groups | 88.63 | 3 | 29.543 | 5.944 | 0.001 |
| | Within Groups | 2450.496 | 493 | 4.971 | | |
| | Total | 2539.126 | 496 | | | |
| FCT | Between Groups | 55.544 | 4 | 13.886 | 1.381 | 0.242 |
| | Within Groups | 1830.027 | 182 | 10.055 | | |
| | Total | 1885.57 | 186 | | | |

### 3.3 Relationship between Height Differences and Terrain Derivatives

The next stage of the analysis considered the relationship between the height differences and first order terrain derivatives, slope and aspect. Figure 8 presents a graphical illustration showing the variabilities. All height differences are observed to occur within slope ranges of $0 - 8°$ and $0 - 10°$ within Lagos and FCT respectively, and within aspect ranges of $0 - 350°$ at both locations. With the exception of the water bodies, which have flat surfaces with a near zero slope and are not used for analysis, the terrain of Lagos is generally gently sloping. The higher slope areas in Lagos State are attributed to the presence of built-up areas and rapidly varying elevations. From Table 6, Pearson's correlation analysis revealed near-zero values of the correlation coefficients (r) between height differences and slope (Lagos, r = 0.014; FCT, r = -0.074). There is no correlation between height differences and aspect at Lagos (r = 0.076) and minimal correlation at FCT (r = 0.258). There is very minimal correlation between the heights and slope in the FCT ($H_{SPOT}$, r = 0.210; $H_{GCP}$, r = 0.212). Generally, this implies very little relationship between the heights and height differences, and slope and aspect. The work of Spaete et al. (2010) showed that there could be a relationship between elevation accuracy and steep slopes for high resolution DEMs. However, this seeming lack of correlation is not conclusive as the analysis suffers from the uneven spatial distribution of GCPs and wide gaps between neighbouring GCPs covering the study area. Figure 8 shows a random occurrence of height differences across areas with varying slope and aspect values in Lagos and the FCT. Figure 8a shows that majority of the slope values in Lagos between $1^0$ and $3^0$ have height differences clustered between -3m and 3m. Fewer points above $3^0$ slope values have height differences with same range. Figure 8b shows the even distribution of the height differences between -3m and 3m across the aspect values ranging from $0^0$ - $350^0$. In the FCT, Figure 8c shows that the height differences between -5m and 5m are clustered around the slope values between $1^0$ and $4^0$. Fewer points above slope value of $4^0$ have height differences within same range. Figure 8d shows the even distribution of the height differences between -5m and 5m across the aspect values ranging from $0^0$ - $350^0$. The evidence does not show any indication that error distribution is a function of slope and/or aspect. However, a denser network of ground for the comparison would be needed for an overarching conclusion.

Table 6: Pearson's correlation coefficient between elevations and terrain derivatives

| Location | | Slope | Aspect |
|---|---|---|---|
| **Lagos** | $H_{SPOT}$ | 0.005 | -0.013 |
| | $H_{GCP}$ | 0.003 | -0.026 |
| | $\Delta H$ | 0.014 | 0.076 |
| **FCT** | $H_{SPOT}$ | 0.210 | 0.025 |
| | $H_{GCP}$ | 0.212 | 0.019 |
| | $\Delta H$ | -0.074 | 0.258 |

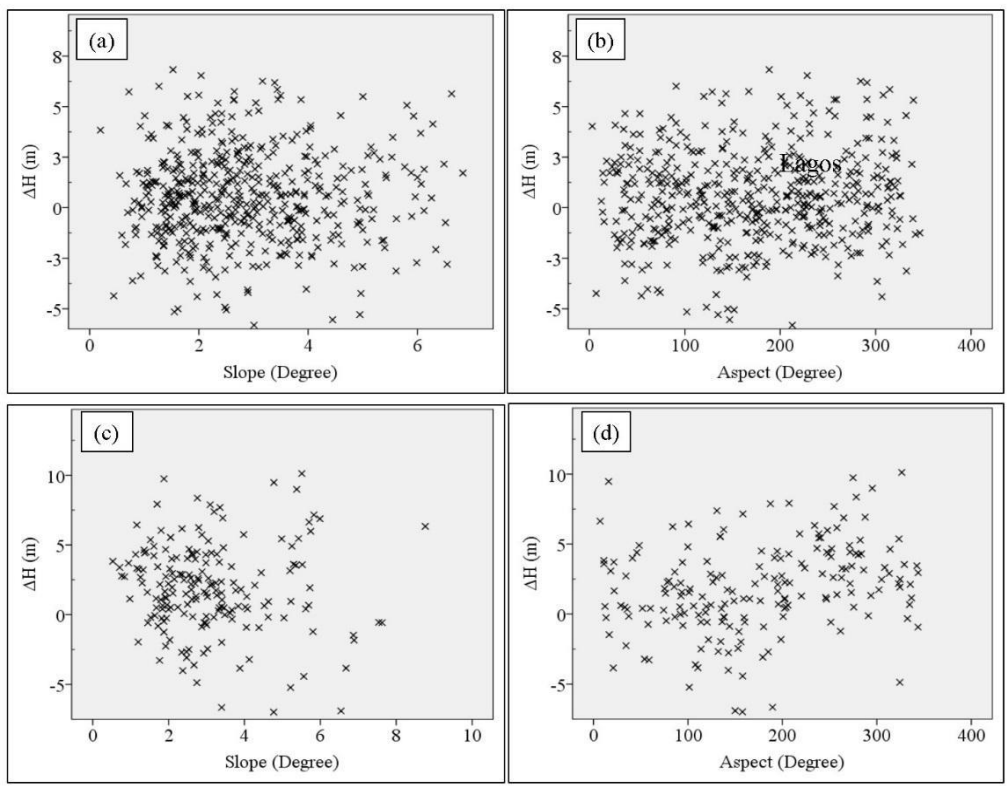

Figure 8: Variability of slope and aspect with height differences, (a) Lagos - slope (b) Lagos – aspect (c) FCT – slope (d) FCT - aspect

## 4. CONCLUSION AND RECOMMENDATIONS

The study presents a quality assessment of SPOT DEM, particularly related to land cover, slope and aspect. From the quality assessment of the DEM across the land cover types for both study sites, it has been revealed that the observed accuracies are within the expected and stated vertical accuracy of 10m for SPOT DEM when the land cover classes are independently considered. Expectedly, and as seen in the SD and RMSE values, the DEM's performance in bare lands is relatively satisfactory. However, the accuracy diminishes in built-up areas and areas of vegetation/forest cover. The heights over the bare lands are the most accurately represented on the DEM with the RMSE values, 2.23m and 2.48m in Lagos and the FCT, respectively, while heights over built-up areas are the least accurate with RMSE values, 2.40m and 4.69m for Lagos and the FCT respectively. The error in overestimating terrain heights is highest in built-up areas, followed by vegetation cover areas. Both study sites have numerous tall structures in urban centres. Lagos is densely populated with buildings both at the residential and commercial land uses. The SPOT DEM's overall accuracy assessment in both study sites has revealed that the RMSE values are 2.33m and 3.69m for Lagos and FCT, respectively. The Pearson's correlation coefficients between the DEM and its derivatives have shown little or no correlation between the DEM and its derivatives, such as aspect and slope. There was a limitation caused by sparse GCP coverage within some parts of the study area. However, the findings present valuable insights into the quality and reliability of the SPOT DEM in the tested areas.

The SPOT DEM's spatial resolution posed significant limitations to effective study and analysis of terrain details. Its coarse 20m resolution is unable to capture elevation differences of fine details such as buildings, trees and other surface objects. Hence, the elevations extracted from the DEM were generalizations of the true surface across a widely varying landscape. This limitation meant specific terrain variability was hidden in the generalizations and could not be analyzed. Moreover, many changes have taken place in the study areas since the DEM was first acquired. Consequently, OSGOF can take advantage of more recent DEMs with finer spatial resolutions such as the 12m TanDEM-X. Also, there is a need for more GCPs per region in future studies to ensure a more holistic assessment. Another recommendation is for OSGOF to integrate the SPOT DEM with other freely available multi-source Global DEMs such as the NASA DEM and ALOS DEM, and digital elevation data from the newly released Global Ecosystem Dynamics Investigation (GEDI) LiDAR. When combined, the complementary characteristics of these multi-source DEMs can yield a fused digital elevation dataset for Nigeria that contains more detailed information than each of the sources. Such a synergized DEM can provide a more hydrologically correct elevation database for environmental modelling.

## ACKNOWLEDGEMENTS

The authors wish to thank the Office of the Surveyor General of the Federation (OSGOF) for provision of the SPOT DEM used in this study. The authors also thank Interspatial

Technologies Ltd, Lagos State Surveyor General's Office and the Federal Capital Development Authority (FCDA), FCT for the Ground Control Points used.

## REFERENCES

Aguilar, F. J., Agüera, F., Aguilar, M. A., and Carvajal, F. (2005). Effects of terrain morphology, sampling density, and interpolation methods on grid DEM accuracy. Photogrammetric Engineering & Remote Sensing, 71(7), 805-816.

Aliyu, R. (2016). Designing for Sustainable Communities: The Abuja Federal Capital Territory of Nigeria. PhD Thesis, De Montfort University, Leicester School of Architecture.

Al-Yami, M.A.M. (2014). Analysis and Visualization of Digital Elevation Data for Catchment Management. (Doctor of Philosophy), University of East Anglia, Norwich.

Arungwa, I.D., Obarafo, E.O. and Okolie, C.J. (2018). Validation of Global Digital Elevation Models in Lagos State, Nigeria. Nigerian Journal of Environmental Sciences and Technology, 2(1), 78 – 88. ISSN (Print): 2616-051X | ISSN (electronic): 2616-0501 https://doi.org/10.36263/nijest.2018.01.0058

A-Xing, Z., Burt, J.E., Smith, M., Rongxun, W., and Jing, G. (2008). The Impact of Neighbourhood Size on Terrain Derivatives and Digital Soil Mapping. In: Zhou Q., Lees B., Tang G. (eds) Advances in Digital Terrain Analysis. Lecture Notes in Geoinformation and Cartography. Springer, Berlin, Heidelberg. https://doi.org/10.1007/978-3-540-77800-4_18

Baudoin, A., Schroeder, M., Valorge, C., Bernard, M., and Rudowski, V. (2004). The HRS-SAP initiative: A scientific assessment of the High-Resolution Stereoscopic instrument on board of SPOT 5 by ISPRS investigators, Proceedings of ISPRS 2004, Istanbul, Turkey, July 12-23, 2004.

Bolstad, P.V., and Stowe, T. (1994). An Evaluation of DEM Accuracy: Elevation, Slope and Aspect. Photogrammetric Engineering & Remote Sensing, 60(11), 7327-7332.

BudgIT. (2018). What we know about Lagos State Finances. https://yourbudgit.com/wp-content/uploads/2018/05/LAGOS-STATE-DATA-BOOK.pdf (Date accessed: December 2nd, 2019).

Chai T., and Draxler R.R. (2014). Root Mean Square Error (RMSE) or Mean Absolute Error (MAE)? - Arguments against avoiding RMSE in the literature. Geosci. Model Dev., 7, 1247–1250. https://doi.org/10.5194/gmd-7-1247-2014

Chigbu, N., Okezie, M., Arungwa, I.D., and Ogba, C. (2019). Comparative analysis of Google Earth Derived Elevation with in-situ Total Station Method for Engineering Constructions. FIG Working Week 2019. Hanoi Vietnam April 22-26, 2019. https://www.fig.net/resources/proceedings/fig_proceedings/fig2019/papers/ts05f/TS05F_chigbu_okezie_et_al_10129.pdf

Crawley, M.J. (2005). Statistics: An Introduction using R, John Wiley and Sons Ltd, West Sussex, England.

Croneborg, L., Saito, K., Matera, M., McKeown, D., and Aardt, J.V. (2015). Digital Elevation Models. A Guidance Note on how Digital Elevation Models are Created and Used – Includes key Definitions, Sample Terms of Reference and how Best to Plan a DEM-Mission. Washington, DC: International Bank for Reconstruction and Development.

Devore, J.L. (2012). Probability and Statistics for Engineering and the Sciences (Eighth ed.): Richard Stratton.

EO Portal Directory (2020). Satellite Missions database – SPOT 5. https://directory.eoportal.org/web/eoportal/satellite-missions/s/spot-5#foot43%29 (Date accessed: May 16, 2020).

Federal Capital Development Authority, FCDA (2020). The Geography of FCT. https://www.fcda.gov.ng/index.php/about-fcda/the-geography-of-abuja (Date accessed: May 18, 2020).

Fisher, P.F., and Tate, N.J. (2006). Causes and consequences of error in digital elevation models. Progress in Physical Geography: Earth and Environment, 30(4), 467–489. https://doi.org/10.1191/0309133306pp492ra

Fisher, P.F. (1991). First experiments in viewshed uncertainty: the accuracy of the viewshed area. Photogrammetric Engineering and Remote Sensing 57, 1321–27.

GISAT (2019). SPOT DEM. http://www.gisat.cz/content/en/products/digital-elevation-model/spot-3d/spot-dem (Date accessed: 2$^{nd}$ December, 2019).

Gorokhovich, Y., and Voustianiouk, A. (2006). Accuracy assessment of the processed SRTM-based elevation data by CGIAR using field data from USA and Thailand and its relation to the terrain characteristics. Remote Sensing of Environment, 104(4), 409-415.

Hengl, T., and Reuter, H.I. (2011). How accurate and usable is GDEM? A statistical assessment of GDEM using Lidar data: In Geomorphometry 2011, edited by T. Hengl, I. S. Evans, J.P. Wilson and M. Gould, 45-48. Redlands, CA, 2011. http://www.geomorphometry.org/HenglReuter2011.

Hirt, C., Filmer, M.S., and Featherstone, W.E. (2010). Comparison and Validation of the recent freely available ASTER-GDEM ver1, SRTM ver4.1 and GEODATA DEM-9S ver3 Digital Elevation Models over Australia. Australian Journal of Earth Sciences, 57(3), 337-347.

IPA (1979). Master Plan for Abuja, the New Federal Capital of Nigeria. Inter. Planning Associates (IPA), 20- 208.

Lee, J., Snyder, P.K., and Fisher, P.F. (1992). Modeling the Effect of Data Errors on Feature Extraction from Digital Elevation Models. Photogrammetric Engineering & Remote Sensing, 58(10), 1461-1467.

Li, Z., and Gruen, A. (2004). Automatic DSM Generation from Linear Array Imagery Data, Proceedings of ISPRS 2004, Istanbul, Turkey, July 12-23, 2004.

Lidberg, W., Nilsson, M., Lundmark, T., and Ågren, A. M. (2017). Evaluating preprocessing methods of digital elevation models for hydrological modelling. Hydrological processes, 31(26), 4660-4668.

Liu, X., Zhang, Z., Peterson, J., and Chandra, S. (2007). The effect of LiDAR data density on DEM accuracy. In Proceedings of the International Congress on Modelling and Simulation (MODSIM07) (1363-1369). Modelling and Simulation Society of Australia and New Zealand Inc..

Mahmoud, I.M., Duker, A., Conrad, C., Thiel, M., and Shaba Ahmad, H. (2016). Analysis of Settlement Expansion and Urban Growth Modelling Using Geoinformation for Assessing Potential Impacts of Urbanization on Climate in Abuja City, Nigeria. Remote Sensing, 8(3), 220. doi:10.3390/rs8030220

Mallo, I.I.Y., and Obasanya, V.G. (2011). Socio-economic effects of Demolishing Squatter Settlements and Illegal Structures in Abuja Metropolis, Federal Capital Territory, Nigeria. Environmental Studies, 7, 10-21.

Massera, S., Favé, P., Gachet, R. and Orsoni, A. (2012). Toward a Global Bundle Adjustment of SPOT-5 HRS Images, Proceedings of the 22nd Congress of ISPRS (International Society of Photogrammetry and Remote Sensing), Melbourne, Australia, Aug. 25 - Sept. 1, 2012, International Archives of the Photogrammetry, Remote Sensing and Spatial Information Sciences, Volume XXXIX-B1, 2012

Maune, D.F. (2011). Digital Elevation Model (DEM) Whitepaper NRCS High Resolution Elevation Data. 1-128.

Miliaresis, C.G. (2008). The Land Cover Impact on the Aspect/Slope Accuracy Dependence of the SRTM-1 Elevation Data for the Humboldt Range. Sensors 2008(8), 3134-3149. doi: 10.3390/s8053134

Mitchell, A. (2005). The ESRI Guide to GIS Analysis, Vol. 2. ESRI Press, Redlands

NTDC (1997). Tourist Attractions in Abuja Destinations. Nigerian Tourism Dev. Corporation (NTDC). Pp 1-8

Nwilo, P.C., Ayodele, E.G. and Okolie, C.J. (2017a). Determination of the Impacts of Landscape Offsets on the 30-metre SRTM DEM through a comparative analysis with Bare-Earth Elevations. FIG Peer Review Journal, 21 pps. ISSN No 2412-916X. http://fig.net/resources/publications/prj/showpeerreviewpaper.asp?pubid=8560

Nwilo, P.C., Okolie, C.J., and Orji, M.J. (2017b). Applications of the 20m SPOT DEM for Geospatial Mapping in Nigeria. Paper presented at the Nigeria Association of Geodesy (NAG), Port-Harcourt, Nigeria.

Nwilo, P.C., Okolie, C.J., Onyegbula, J.C., Abolaji, O.E., Orji, M.J., Daramola, O.E. and Arungwa, I.D. (2020). Vertical Accuracy Assessment of 20-metre SPOT DEM using Ground Control Points from Lagos and Abuja, Nigeria. Manuscript under review.

Ojigi, L.M. (2006). Analysis of Spatial Variations of Abuja Land Use and Land Cover from Image Classification Algorithms. ISPRS Commission VII Mid-Term Symposium. Theme: Remote Sensing: From Pixel to Processes. 8 – 11th May 2006, Enschede, The Netherlands.

Olusina, J.O., Okolie, C.J., and Emmanuel, I.E. (2018). Quality Assessment of Terrain Derivatives from 30m Global DEMs: SRTM v3.0 And ASTER v2. Paper Presented at the 13th University of Lagos Research Conference and Fair, University of Lagos Nigeria.

Olusina, J.O., and Okolie, C.J., (2018). Visualisation of Uncertainty in 30m Resolution Global Digital Elevation Models: SRTM v3. 0 and ASTER v2. Nigerian Journal of Technological Development, 15(3), 77-83.

Pidwirny, M. (2006). Atmospheric Effects on Incoming Solar Radiation. Fundamentals of Physical Geography, 2nd Edition. http://www.physicalgeography.net/fundamentals/7f.html

Podobnikar, T. (2008). Methods for visual quality assessment of a digital terrain model. S.A.P.I.EN.S. 1(2), 1-10. http://sapiens.revues.org/index738.html.

Qiming, Z., and Xuejun, L. (2008) Assessing Uncertainties in Derived Slope and Aspect from a Grid DEM. In: Zhou Q., Lees B., Tang G. (eds) Advances in Digital Terrain Analysis. Lecture Notes in Geoinformation and Cartography. Springer, Berlin, Heidelberg. https://doi.org/10.1007/978-3-540-77800-4_15

Ravibabu, M.V., and Jain, K. (2008). Digital Elevation Model Accuracy Aspects. Journal of Applied Sciences, 8, 134-139.

Reinartz, P., Lehner, M., Müller, R., and Schroeder, M. (2004). Accuracy Analysis from DEM and Orthoimages derived from SPOT HRS Stereo Data without using GCP, Proceedings of ISPRS 2004, Istanbul, Turkey, July 12-23, 2004.

Rexer, M., and Hirt, C. (2014). Comparison of Free High-Resolution Digital Elevation Datasets (ASTER GDEM2, SRTM v2.1/v4.1) and Validation against Accurate Heights from the Australian National Gravity Database; Australian Journal of Earth Sciences, 1-15, doi: 10.1080/08120099.2014.884983, 2014.

Rosengren, M., and Willén, E. (2004). Multiresolution SPOT-5 data for boreal forest monitoring. Paper presented at the Proceedings of ISPRS Congress, Istanbul, July 12.

Schlaffer, S., Matgen, P., Hollaus, M., and Wagner, W. (2015). Flood detection from multi-temporal SAR data using harmonic analysis and change detection. In International Journal of Applied Earth Observation and Geoinformation 38, 15–24.

Spaete, L.P., Glenn, N.F., Derryberry, D.R., Sankey, T.T., Mitchell, J.J., and Hardegree, S.P. (2010). Vegetation and slope effects on accuracy of a LiDAR-derived DEM in the sagebrush steppe. Remote Sensing Letters, 2(4), 317–326. doi:10.1080/01431161.2010.51526

Spiegel, M. R., and Stephens, L. J. (1999). Theory and Problems of Statistics (Third Edition ed.).

Tsyganskaya, V., Martinisb, S., Tweleb, A., Caob, W., Schmittb, A., Marzahna, P., and Ludwiga, R. (2016). A Fuzzy Logic-Based Approach for the Detection of Flooded Vegetation by means of Synthetic Aperture Radar Data. The International Archives of the Photogrammetry, Remote Sensing and Spatial Information Sciences, XLI-B7, 371-378. XXIII ISPRS Congress, 12–19 July 2016, Prague, Czech Republic. doi:10.5194/isprsarchives-XLI-B7-371-2016.

Tighe M.L., and Chamberlain, D. (2009). Accuracy Comparison of the SRTM, ASTER, NED, NEXTMAP USA Digital Terrain Model over several USA Study Sites. ASPRS/MAPPS 2009 Fall Conference. November 16-19, 2009 San Antonio, Texas.

Townsend, P.A (2001). Mapping seasonal flooding in forested wetlands using multi-temporal radarsat SAR. In Photogrammetric Engineering and Remote Sensing 67(7), 857–864.

Wang, J., and Maduako, I.N. (2018). Spatio-temporal urban growth dynamics of Lagos Metropolitan Region of Nigeria based on Hybrid methods for LULC modeling and prediction. European Journal of Remote Sensing, 51(1), 251–265. doi:10.1080/22797254.2017.1419831

Wang, P. (198). Applying Two Dimensional Kalman Filtering for Digital Terrain Modelling. In D. Fritsch, M. Englich & M. Sester, eds, 'IAPRS', Vol. 32/4, ISPRS Commission IV Symposium on GIS - Between Visions and Applications, pp, 649-656. Stuttgart, Germany. IAPRS, Vol. 32, Part 4 "GIS-Between Visions and Applications", Stuttgart, 1998

Yastikli N., Kocak G., and Buyuksalih G. (2006). Accuracy and Morphological Analyses of GTOPO30 and SRTM X -C Band DEMs in the test area Istanbul. Paper presented at ISPRS Workshop on Topographic Mapping from Space (with Special Emphasis on Small Satellites), Ankara, Turkey, February 14-16.

Zhou, Q., and Liu, X. (2004). Analysis of errors of derived slope and aspect related to DEM data properties. Computers & Geosciences, 30(4), 369-378.